\documentclass[aps,twocolumn,preprintnumbers,amsmath,amssymb,showpacs]{revtex4}

\def\be{\begin{equation}}
\def\ee{\end{equation}}

\def\be{\begin{equation}}
\def\ee{\end{equation}}

\def\be{\begin{equation}}
\def\ee{\end{equation}}

\usepackage[dvipdfmx]{graphicx}
\usepackage{here,braket,booktabs}
\usepackage{comment}
\usepackage{epsf}
\usepackage{amsmath}
\usepackage{graphics}
\usepackage{amsfonts}
\usepackage{amssymb}
\usepackage{latexsym}
\usepackage{color}
\input{colordvi.tex}
\usepackage{feynmp}

\begin{document}

\title{Role of spacetime boundaries in Einstein's other gravity}

\author{Naritaka Oshita$^{1,2}$ and Yi-Peng Wu$^{1}$}
\affiliation{
  ${}^1$Research Center for the Early Universe (RESCEU), Graduate School
  of Science,\\ The University of Tokyo, Tokyo 113-0033, Japan
}
\affiliation{
  ${}^2$Department of Physics, Graduate School of Science,\\ The University of Tokyo, Tokyo 113-0033, Japan
}

\preprint{RESCEU-7/17}

\begin{abstract}
Einstein's vierbein formulation of general relativity based on the notion of distant parallelism (teleparallelism) naturally
introduces a covariant surface term in addition to the Einstein-Hilbert action. We investigate the action principle in teleparallelism
with the existence of spacetime boundaries and find that the covariant surface term exactly eliminates all the unwanted surface terms reside
in the metric formulation of general relativity, as the role of a Gibbons-Hawking-York (GHY) term. 
The identity of such a covariant GHY term is further confirmed by the recovery of the correct black hole entropy from 
the free energy due to the spacetime boundary.
These results indicate that the vierbein formulation of gravity generally exhibits a well-posed action principle and 
readily admits the path integral approach to quantization. 
\end{abstract}


\maketitle

\section{Introduction}
The Einstein-Hilbert action appears to be insufficient to reproduce 
the field equations of general relativity (GR) when boundaries of the spacetime play a fundamental role in the variation
with respect to the dynamical variable of gravity. 
To remedy the action principle of GR, one shall include surface terms contributed by 
the extrinsic curvature of the boundaries embedded in spacetime~\cite{York:1972sj,Gibbons:1976ue},
provided that dynamical degrees of freedom are fixed on the boudary surfaces.
These surface terms in addition to the Einstein-Hilbert action are not covariant 
but they are essential for the path-integral approach to quantum gravity \cite{Gibbons:1976ue,Hawking:1978},
since the final result must be independent of any intermediate hypersurface along the path of integration.

However, as additional surface terms must be supplied in the action principle with spacetime boundaries, 
one may wonder if there exsits a unified Lagrangian that always results in the correct field equations of GR,
or, if more importantly, an appropriate action constructed by a covariant formulation.
In this work, we show that both requirements are indeed realized in Einstein's alternative description of GR 
when adopting the notion of distant parallelism \cite{Einstein:1928,Cartan}.

The concept of distant parallelism (or teleparallelism) is proposed by Einstein
for a unified theory of gravitation and electromagnetic fields \cite{Einstein:1928}.
To have a clear notion of parallelism for two vectors separated by a finite distance, which seems to be insufficient
in GR, one naturally  
postulates the existence of the vierbein field $e_A^{\;\mu}$ at every (four-dimensional) spacetime point
and imposes the rule of absolute parallel transport as $\nabla_\mu e_A^{\;\;\nu} = 0$, where capital Latin indices stand for
components in a local tangent frame. This special parallel transport is realized by the so-called
Weitzenb\"{o}ck connection \cite{Weitzenbock:1923}
$\Delta^\lambda_{\;\;\nu\mu} = e_A^{\;\;\lambda}\partial_\mu e^A_{\;\;\nu}$ whose corresponding
Riemannian tensor vanishes identically \cite{Einstein:1928,Aldrovandi:2012}.
 
One can construct a covariant theory in teleparallelism based on the torsion tensor
$T^\lambda_{\;\;\mu\nu} \equiv
\Delta^\lambda_{\;\;\nu\mu} -\Delta^\lambda_{\;\;\mu\nu}$, which contains only first derivatives of the vierbein field.
By the subtraction of the Weitzenb\"{o}ck connection we have removed in the torsion tensor all the contribution from the
metric $g_{\mu\nu} = \eta_{AB} e^A_{\;\mu}e^B_{\;\nu}$, where $\eta_{AB} = \text{diag}(-1,1,1,1)$. 
However, among all the possible combinations of the quadratic torsion tensors only one certain class of choices is found to result in
symmetric field equations, which was firstly developed by Einstein \cite{Einstein:1929} (see also \cite{Cho:1975dh,Hayashi:1979qx,Maluf:1994}). 
We denote this class of theory by the action
\begin{equation}
\label{action:T}
S_{\rm T} = \frac{1}{16\pi G}\int d^4 x \; e \; T,
\end{equation}
where $e \equiv \mathrm{det}[e^A_{\;\mu}] = \sqrt{-g}$ and $T$ is the unique torsion scalar defined as \eqref{Lagrangian:T}.
To the purpose of the current study, it is more convenient to reformulate the torsion scalar as
\begin{equation}
\label{T-R relation}
T = \bar{R} + 2\bar{\nabla}_\mu T^{\alpha\mu}_{\;\;\;\;\alpha},
\end{equation}
where $\bar{R}$ is the Ricci scalar and 
$\bar{\nabla}_\mu$ is defined as the covariant derivative with respect to the
Levi-Civita connection used in GR \cite{Wu:2016dkt}. From the formulation \eqref{T-R relation} we easily recognize that in the action of the
teleparallel theory \eqref{action:T} the only difference to the Einstein-Hilbert action
\begin{equation}
\label{action:GR}
S_{\rm GR} = \frac{1}{16\pi G}\int d^4 x \; \sqrt{-g} \; \bar{R},
\end{equation}
is given by a surface term.
Thus the variation of the action \eqref{action:T} with respect to the vierbein field solely leads to 
equivalent field equations as those obtained from the action principle of GR,
provided that one can ignore any contribution from the spacetime boundaries.  

In this work, we investigate the contribution of surface terms in the action 
principle of the teleparallel theory \eqref{action:T}, assuming that the spacetime has a boundary.
We first prove that the surface term purely due to torsion in \eqref{T-R relation} exactly 
eliminates all the other surface terms reside in $\bar{R}$ once the metric on the boundary is fixed. 
Such a result coincides with the purpose of introducing a surface term by Gibbons and Hawking \cite{Gibbons:1976ue}
to remove the second derivatives in the Einstein-Hilbert action, as
York \cite{York:1972sj,York:1986} found that the non-vanished surface term in GR  
modifies the dynamics of gravitation.

We then study the entropy contributed by the covariant surface term in a spacetime with an event horizon.
To draw a comparison to the result given by the path integral approach \cite{Gibbons:1976ue,Hawking:1978},
we rotate the time axes into complex directions (Wick rotations) to obtain the Euclidean torsion scalar,
following the spacetime decomposition for the teleparallel theory developed in \cite{Wu:2011kh,Wu:2016dkt}.
We apply the regularization used in \cite{Hawking:1995fd} to obtain a finite energy from the spacetime boundary and
confirm that the covariant surface term contributes a correct amount of black hole entropy, as a consistent finding with \cite{Gibbons:1976ue,Hawking:1978}.

One can read this paper from two equivalent perspectives.
The first one is to stay in Riemannian geometry where $\bar{R}$ is the curvature scalar of the spacetime.
In this case $T^\lambda_{\;\;\mu\nu} =e_A^{\;\;\lambda}(\partial_\mu e^A_{\;\;\nu}-\partial_\nu e^A_{\;\;\mu})$ is a tensor
defined by the vierbein field but has no further geometrical meaning.
This is the original approach considered by Einstein \cite{Einstein:1929}.
The second perspective is to stay in Weitzenb\"{o}ck geometry where $\Delta^\lambda_{\;\;\mu\nu}$ defines the spacetime connection
and the curvature of the spacetime is always zero. In this case $\bar{R}$ is a scalar defined by the metric but has no further geometrical meaning.
This is the approach where teleparallelism becomes a pure theory for gravitation 
\cite{Aldrovandi:2012,Cho:1975dh,Hayashi:1979qx,Maluf:1994,Wu:2016dkt}.

\section{The Gibbons-Hawking-York term}

We shall consider a non-null spacetime boundary $\partial \mathcal{M} : f(x^\mu) = \rm constant$, which introduces a family  of unit normal 
$n_\mu\equiv \partial_\mu f/\sqrt{\epsilon g^{\alpha\beta}\partial_\alpha f\partial_\beta f}$.
The non-null condition for the boundary reads $n^\mu n_\mu = \epsilon = \pm 1$ 
where $\epsilon= +1$ and $\epsilon = -1$ correspond to a timelike and spacelike hypersurface respectively.

The action principle for GR is well-known:
\begin{align}\label{variation:GR1}
\delta S_{\rm GR} &= \frac{1}{16\pi G} \int_{\cal M} d^4 x \; \sqrt{-g} \; G_{\mu\nu} \delta g^{\mu\nu} \\\nonumber
&+\frac{ \epsilon}{16\pi G}\oint_{\partial \mathcal{M}}d^3\Omega \sqrt{h} \; n^\mu 
g^{\alpha\beta}(\partial_\alpha\delta g_{\mu\beta} - \partial_\mu \delta g_{\alpha\beta}),
\end{align}
where $h$ is the absolute value of $\mathrm{det}[h_{\mu\nu}]$ and 
\begin{equation}\label{metric:induced}
h_{\mu\nu} = g_{\mu\nu} - \epsilon n_\mu n_\nu,
\end{equation}
is the induced metric on the boundary surface.

If the metric on the boundary is fixed such that $\delta g_{\mu\nu}\vert_{\partial \mathcal{M}} = 0$,
then one can eliminate the contribution from the second term on the right-hand-side of \eqref{variation:GR1} by
imposing the Gibbons-Hawking-York (GHY) term \cite{York:1972sj,Gibbons:1976ue}
\begin{equation}\label{YGH-term}
S_{\rm GHY} = \frac{\epsilon}{8\pi G}\oint_{\partial \mathcal{M}}d^3\Omega \sqrt{h} \bar{K},
\end{equation}
where $\bar{K} \equiv \bar{\nabla}_\mu n^\mu$ is the trace of the extrinsic curvature of the boundary defined in Riemannian spacetime.
(Note that the extrinsic curvature $K_{\mu\nu} = \nabla_\mu n_\nu$ is not symmetric when the spacetime contains torsion.)

According to the boundary condition $\delta g_{\mu\nu} = 0$, we have $h^{\alpha\beta}\partial_\alpha \delta g_{\mu\beta} = 0$
where the variation of $S_{\rm GHY} $ only survives in the normal direction to the surface. Therefore one can derive
\begin{equation}
\delta S_{\rm GHY} = \frac{\epsilon}{16\pi G}\oint_{\partial \mathcal{M}}d^3\Omega \sqrt{h} n^\mu h^{\alpha\beta}\partial_\mu \delta g_{\alpha\beta}.
\end{equation}
On the other hand, we can replace $g^{\alpha\beta}$ in the surface term of \eqref{variation:GR1} 
by the induced metric according to \eqref{metric:induced}, which gives
\begin{align}
\label{variation:GHY}
\delta S_{\rm GR} + \delta S_{\rm GHY} = \frac{1}{16\pi G} \int_{\cal M} d^4 x \; \sqrt{-g} \; G_{\mu\nu} \delta g^{\mu\nu}.
\end{align}
As a result, with the existence of the GHY term a proper field equation can be derived from the action principle in a spacetime with a boundary.

\section{The action principle in teleparallel theory}

We now prove that the GHY term is naturally embedded in the theory \eqref{action:T}.
The first step is to rewrite the well-known results of GR in terms of the vierbein field. Since
$\delta g_{\mu\nu} = \eta_{AB} (\delta e^A_{\;\mu} e^{B}_{\;\nu} +  e^A_{\;\mu} \delta e^B_{\;\nu})$, 
one finds $\delta e^A_{\;\mu} e^{B}_{\;\nu} =-  e^A_{\;\mu} \delta e^B_{\;\nu}$ is always satisfied for any
$e^A_{\mu}$ when $\delta g_{\mu\nu}$ vanishes.
As a result the imposed boundary condition indicates
\begin{equation}
\label{boundary_condition}
\delta g_{\mu\nu}\arrowvert_{\partial \mathcal{M}} = 0 \; \Leftrightarrow \; \delta e^A_{\;\;\mu}\arrowvert_{\partial \mathcal{M}} = 0.
\end{equation}
Note that only ten components in $e^A_{\;\mu}$ are fixed by this boundary condition, while the rest six
components are attributed to an arbitrary local Lorentz transfromation in the tangent frame.

On the other hand, the induced metric \eqref{metric:induced} is given by 
$h_{\mu\nu}  = (\eta_{AB} - \epsilon n_A n_B)e^A_{\;\mu} e^B_{\;\nu}$, where $n_A = n_\mu e_A^{\;\mu}$
is the component of a unit normal vector in a local tangent frame.
After some simple rearrangement we can decompose the vierbein field on the boundary as
\begin{equation}
\label{vierbein_on_boundary}
e_A^{\; \mu} = h_A^{\;\mu} + \epsilon n_A n^\mu.
\end{equation}
Given that the vierbein field is fixed on the boundary, the derivative tangent to the surface must vanish, 
that is $h_A^\mu\partial_\mu\delta e^B_\nu = 0$. Therefore we may rewrite \eqref{variation:GR1} as
\begin{align}
\label{variation:GR2}
\delta S_{\rm GR} &= \frac{1}{16\pi G}\int_{\cal M} d^4 x \; \sqrt{-g} \; G_{\mu\nu} \delta g^{\mu\nu} \\\nonumber
&+ \frac{\epsilon}{8\pi G}\oint_{\partial \mathcal{M}}d^3\Omega \sqrt{h} \; n^\mu 
\left[e_A^\alpha\partial_\alpha\delta e^A_{\;\mu} - e_A^{\;\alpha} \partial_\mu\delta e^A_{\;\alpha}\right].
\end{align}

Let us now check the contribution from the torsion induced surface term in \eqref{T-R relation},
which reads
\begin{align}\label{torsion_surface}
\int_\mathcal{M} d^4x \;  e \;\bar{\nabla}_\mu T^{\alpha\mu}_{\;\;\;\;\alpha} = 
\epsilon \oint_{\partial \mathcal{M}}d^3\Omega \sqrt{h} \; n_\mu T^{\alpha\mu}_{\;\;\;\;\alpha},
\end{align}
As $\delta n_\mu = \delta g^{\mu\nu} =  \delta e^A_{\;\;\mu} = 0$ on the boundary,
the non-vanished contribution from variation is
\begin{align}\label{variation:torsion_surface}
\delta \left[ n_\mu T^{\alpha\mu}_{\;\;\;\;\alpha}\right]=   n^\mu 
\left[ e_A^{\;\alpha} \partial_\mu\delta e^A_{\;\alpha}- e_A^\alpha\partial_\alpha\delta e^A_{\;\mu}  \right].
\end{align}
We can combine the results from \eqref{variation:GR2} and \eqref{variation:torsion_surface} to obtain
the variation of the teleparallel theory as
\begin{equation}
\label{variation:T}
\delta S_{\rm T} = \frac{1}{16\pi G} \int_\mathcal{M} d^4 x \; \sqrt{-g} \; G_{\mu\nu} \delta g^{\mu\nu},
\end{equation}
where the surface term due to the Ricci scalar $\bar{R}$ is exactly cancelled out by the 
torsion induced surface term. This result is the identical to what happened in GR when a
surface term of \eqref{YGH-term} is added into the action \eqref{action:GR}.

\section{Spacetime free energy from the covariant boundary term}
We have shown that the torsion induced surface term \eqref{torsion_surface} acts as a GHY term in the action principle.
The necessity of a GHY term is firstly supported by its contribution to 
the ``thermodynamic potential'' (or namely the free energy) of the homotopically disconnected metrics 
where the spacetime has event horizons (such as the Schwarzschild solution for black holes)~\cite{Gibbons:1976ue,Hawking:1978}. 
Here we examine the free energy contributed by the covariant surface term in the theory \eqref{action:T}, that is
\begin{eqnarray}\label{entropy:torsion_surface}
S_{c}\equiv \frac{1}{8 \pi G} \int_{\mathcal M} d^4 x\; \tilde{e}\; \bar{\nabla}_{\mu} T^{\nu\mu}_{\;\;\;\; \nu}.
\end{eqnarray}

We apply the Wick rotation to both the time coordinate in spacetime and in the local tangent frame to 
obtain the Euclidean metric $\tilde{g}_{\mu\nu} = \tilde{\eta}_{AB} \tilde{e}^A_{\;\mu}\tilde{e}^B_{\;\nu}$, 
where $\tilde{e}^A_{\;\mu}$ is the Euclidean vierbein field and $\tilde{\eta}_{AB} = \text{diag}(1,1,1,1)$. 
It is convenient to start with the general form \cite{Wu:2011kh,Wu:2016dkt}
\begin{eqnarray}
\label{ADM-vierbein 0th}\nonumber
\tilde{e}^0_{\;\mu}=(N,{\bf 0})\;\; &,&\;\;\;
\tilde{e}^a_{\;\mu}=(\tilde{N}^a,h^a_{\,\,i})\;,\\
\tilde{e}^{\,\,\mu}_{0}=(1/N,-\tilde{N}^i /N)\;\; &,&\;\;
\tilde{e}^{\,\,\mu}_{a}=(0,h_a^{\,\,i})\;,
\end{eqnarray}
which gives rise to the Euclidean metric in terms of the Arnowitt-Deser-Misner (ADM) formalism \cite{Arnowitt:1959ah,Oshita:2016oqn}:
\begin{align}
d\tilde{s}^2 = N^2d\tilde{t}^2 + h_{ij} \left(dx^i + \tilde{N}^id \tilde{t}\right)\left(dx^j +\tilde{N}^j d\tilde{t}\right),
\end{align}
where $h_{ij} = \eta_{ab}h^a_{\,\,i}h^b_{\,\,j}$ is the induced metric on the spacelike 3-surface, and $h^a_{\,\,i}$ is namely the
induced vierbein (or the dreibein field) \cite{Wu:2016dkt,Indices}.
Here $\tilde{t} = i t$ is the Euclidean time and $\tilde{N}^i = -iN^i$ where $N^i$ is the shift vector without Wick rotation.
The Euclidean torsion scalar in this spacetime decomposition reads \cite{Wu:2016dkt}
\begin{equation}
\tilde{T} =  \tilde{R}^3 + \tilde{K}^2 - \tilde{K}^{ij}\tilde{K}_{ij}  + \frac{2}{N} \bar{D}_i (h^{ij} N T^{k}_{\,\, j k}),
\end{equation}
where $\tilde{K}_{ij} = -i \bar{K}_{ij}$ is the (Euclidean) extrinsic curvature defined in GR, 
$\bar{D}_i$ is the 3-covariant derivative with respect to the 3-Levi-Civita connection,
and $T^{i}_{\,\, j k} = h^i_{\; a}(\partial_k h^a_{\; j} - \partial_j h^a_{\; k})$ can be recognized as the 3-torsion
in view of the Weitzenb\"{o}ck geometry (which is unchanged under Wick rotation).

\begin{figure}[H]
	\begin{center}
		\includegraphics[keepaspectratio=true,height=50mm]{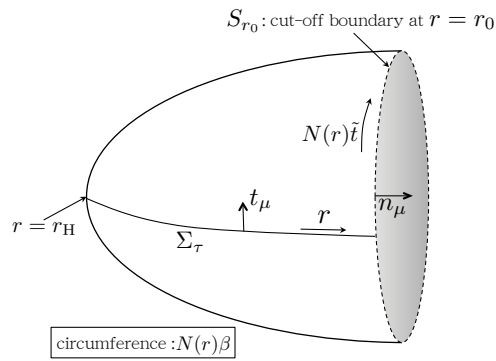}
	\end{center}
	\caption{The asymptotically flat Euclidean manifold ${\mathcal M}$ with a horizon at $r = r_{\text{H}}$
		and a cut-off boundary $S_{r_0}$ is imposed at $r = r_0$.
	}%
	\label{0509fig}
\end{figure}

We now consider a spherically symmetric, stationary, and asymptotically flat Euclidean spacetime $\mathcal M$,
where the vierbein field and metric of ${\mathcal M}$ take the forms
\begin{align}
\tilde{e}^{A}_{\,\,\mu} &= \text{diag} \left(\sqrt{f(r)}, \frac{1}{\sqrt{f(r)}},r,r \sin{\theta} \right),\\
ds^2 &= f(r) d\tilde{t}^2 + f^{-1} (r) dr^2 +r^2 d\Omega_2^2,
\label{metric:spherical_stationary}
\end{align}
respectively.
In this case $N = \sqrt{f(r)}$, $\tilde{N}^a = \tilde{N}^i h^a_{\;i} =0$, and 
the induced metric on the spacelike hypersurface $\Sigma_{\tau}$ fixed at a Euclidean time
$\tilde{t} = \tau$ can be derived as $h_{\mu \nu} =g_{\mu \nu} -t_{\mu} t_{\nu} = \text{diag} ( 0,1/f,r^2,r^2 \sin^{2}\theta )$,
where $t_{\mu}= \text{diag}(\sqrt{f},0 ,0,0)$ is the unit normal vector of $\Sigma_{\tau}$.
The dreibein field is $h^a_{\,\,i}= \text{diag} ( 0,1/\sqrt{f} ,r,r \sin\theta )$.
We shall impose a horizon at $r = r_H$, such that $f (r_H) = 0$, where the horizon merges to a conical singularity 
of the Euclidean spacetime as seen by Fig. \ref{0509fig}.

Let us introduce a boundary surface $S_{r_0}$ at $r =r_0$ where 
the normal vector of $S_{r_0}$ is given by
\begin{eqnarray}
n_{\mu} = (0, 1/\sqrt{f (r)}, 0,0).
\label{050502}
\end{eqnarray}
One can regard the time parameter $\tau$ as an angular variable with the periodicity $\tau = \beta$,
where $1/\beta \equiv |f'(r_H)|/4 \pi$ is the definition of the Hawking temperature \cite{Padmanabhan:2009vy}.
The integration \eqref{entropy:torsion_surface} reads
\begin{align}\label{entropy:torsio_surface2} \nonumber
S_{c}^{\text{(E)}} 
&= -\frac{1}{8 \pi G} \displaystyle \int_{0}^{\beta} d \tau \int_{S_{r_0}} d^2x  \sqrt{\sigma} \;
N n_{i} h^{ij}\tilde{T}^{\nu}_{\;\;j\nu},
\\
&= -\frac{\beta}{8 \pi G} \displaystyle \int_{S_{r_0}} d^2x  \sqrt{\sigma} \;
N n^{r} \tilde{T}^{\nu}_{\;\;r\nu},  \\\nonumber
& = -\frac{\beta}{8 \pi G} \displaystyle \int_{S_{r_0}} d^2x \sqrt{\sigma} \;
f(r_{0}) \left( \frac{1}{2} \frac{f^\prime(r_{0})}{f (r_{0})} +\frac{2}{r_0} \right).
\end{align}
where $S_c^{\text{(E)}} \equiv -i S_c$ and $\sqrt{\sigma}$ is the Jacobian of the two-sphere with $\sigma \equiv \text{det} [\sigma_{\mu \nu}]$ and
$\sigma_{\mu \nu} \equiv h_{\mu \nu} - n_{\mu} n_{\nu}$.
Note that $\beta N$ is the length of the circumference of the spacetime $\cal M$.

For an asymptotically flat spacetime the physical action has to be regularized by a reference background 
defined by the flat spacetime \cite{Hawking:1995fd}. To this aim we also introduce a reference term
 \begin{align}
 S_{c 0}^{\text{(E)}} \equiv -\frac{1}{8 \pi G} \int_{\mathcal M} d^4 x\; \tilde{e}\; \bar{\nabla}_{\mu} [\tilde{T}^{\nu\mu}_{\;\;\;\; \nu}]_0,
 \end{align}
 where $[\tilde{T}^{\nu\mu}_{\;\;\;\; \nu}]_0$ is the torsion tensor given by a flat background with the same 
 circumference $\beta N$ as $\cal M$. The vierbein field for the reference background can be chosen as
 $[\tilde{e}^{A}_{\,\,\mu}]_0 = \text{diag} (\sqrt{f(r)}, 1,r,r \sin{\theta} )$.

Let us take the example $f (r) = 1-2GM/r$, where the metric \eqref{metric:spherical_stationary}
is identified with the Schwarzschild solution with a black hole mass $M$.
In this case the integration \eqref{entropy:torsio_surface2} becomes
\begin{eqnarray}\label{Sc:unregularized1}
S_{c}^{\text{(E)}} =-4 \pi M (2r_0-3GM),
\end{eqnarray}
where the first term diverges as $r_0 \to \infty$.
However, the physical result is obtained from the regularization
\begin{align}
\label{result:free_energy}
S_{c}^{\text{(E)}} - S_{c 0}^{\text{(E)}} = 4 \pi G M^2,
\end{align} 
where one can check that $S_{c 0}^{\text{(E)}} \to  -4 \pi M (2 r_0 - 2GM)$ as $r_0 \gg 2GM$ 
for the integration over the boundary $S_{r_0}$.

Keeping in mind that the energy of the Schwarzschild metric 
$\langle E\rangle = M$ \cite{Hawking:1978} and that the inversed Hawking temperature
$\beta = 8\pi GM$, we can reproduce the correct Bekenstein entropy \cite{Bekenstein:1973ur}
\begin{align}
S= \beta M - F = 4\pi GM = \frac{A}{4G},
\end{align}
if we recognize the final result \eqref{result:free_energy} as the free energy $F$ of the system.
Here $A\equiv 4 \pi (2GM)^2$ is the area of the event horizon.
Given that the GHY term is responsible to the spacetime free energy in the path integral approach \cite{Gibbons:1976ue,Hawking:1978},
once agian we confirm that the covariant surface term $S_c$ made by the vierbein field plays the role of the GHY term 
in the teleparallel formulation of GR~\eqref{action:T}.

\section{Conclusion}

In this work we have revisited the so-called ``teleparallel equivalent of general relativity'' by further considering the existence of spacetime boundaries.
We found that the boundary condition $\delta e^A_{\;\;\mu} =0$ is enough to make sure 
the action to be stationary, given that the unique Lagrangian formulated by the quadratic torsion tensor \cite{Einstein:1929}
\begin{align}
\label{Lagrangian:T}
T= -\frac{1}{4} T_\lambda^{\;\;\mu\nu}T^\lambda_{\;\;\mu\nu} + \frac{1}{2}T_{\quad\lambda}^{\mu\nu}T^\lambda_{\;\;\mu\nu}
+ T^\lambda_{\;\;\lambda\alpha}T^{\rho\;\;\alpha}_{\;\;\rho},
\end{align}
contains only first derivatives of the vierbein field.
One can rewrite the torsion scalar in terms of a Ricci scalar $\bar{R}$ and a covariant surface term as seen by \eqref{T-R relation}.
This reformulation of $T$ is realized by a decomposition of the Weitzenb\"{o}ck connection 
$\Delta^\lambda_{\;\;\nu\mu} = e_A^{\;\;\lambda}\partial_\mu e^A_{\;\;\nu}$ into a Levi-Civita connection and a contorsion tensor
$K^\lambda_{\;\mu\nu} = \frac{1}{2}(T^{\;\;\lambda}_{\mu\;\;\nu} + 
T^{\;\;\lambda}_{\nu\;\;\mu} - T^\lambda_{\;\;\mu\nu})$ \cite{Aldrovandi:2012}.

The surface term in the Einstein-Hilbert action 
 is known to play a fundamental role in the action principle of GR \cite{York:1972sj}.
Therefore, to result in a consistent variation of the teleparallel theory \eqref{action:T} with both the 
representation \eqref{T-R relation} and \eqref{Lagrangian:T}, the covariant surface term $S_c$ must 
exactly cancel the surface term residing in $\bar{R}$, as what is done by a GHY term \cite{Gibbons:1976ue,York:1986}.
Combining the results \eqref{variation:GHY} and \eqref{variation:T}, we have verified that
\begin{equation}
\delta S_T = \delta S_{\rm GR} + \delta S_{\rm GHY},
\end{equation}
where, to our best knowledge, this is the first identification of $\delta S_c$ with $\delta S_{\rm GHY}$ in the action principle,
despite the contribution of $S_c$ to the gravitational stress-energy has been discussed in some literature \cite{Krssak:2015lba,Krssak:2015rqa}.

We shall remark that the boundary condition $\delta e^A_{\;\;\mu} =0$ is sufficient to give a well-posed action principle even to
theories based on the generalized scalar-tensor formulation of the torsion scalar $T$ 
(which includes the $f(T)$ gravity as a special case)  \cite{Wu:2016dkt,Sotiriou:2010mv,Geng:2011aj,Jarv:2015odu}.
There is no need to impose any counter-boundary terms 
due to the fact that the Lagrangian of the teleparallel scalar-tensor theory contains only the first derivatives of the
vierbein field \cite{Li:2010cg}.
 On the other hand, to ensure a stationary action for the scalar-tensor theory with respect to $\bar{R}$, including the
$f(\bar{R})$ gravity, one must not only introduce a modified GHY term and impose the boundary condition $\delta g_{\mu\nu} =0$,
but also fix the additional scalar degree of freedom on the boundary surface \cite{Dyer:2008hb}.

Finally, we emphasize that the covariant GHY term $S_c$ is only realized by
treating the vierbein field as the true dynamical degrees of freedom in gravitation,
such that one can construct a
covariant Lagrangian not only based on the metric but also the specific tensor
$T^\lambda_{\;\;\mu\nu} =e_A^{\;\;\lambda}(\partial_\mu e^A_{\;\;\nu}-\partial_\nu e^A_{\;\;\mu})$ in the Riemannian spacetime
(that has no torsion)~\cite{Einstein:1928}. The tensor $T^\lambda_{\;\;\mu\nu}$ defined by the vierbein field and its first derivative coincides with the
torsion tensor in the Weitzenb\"{o}ck spacetime (that has no curvature).
We believe that the well-posed action principle in the vierbein formulation of GR shows one significant benefit 
to postulate the existence of the vierbein field, aside from the importance of the vierbein field
in a unified theory of gravitation and electromagnetic field \cite{Nelson:1977qj,Nelson:1978ex},
or in the construction of quantum gravity \cite{Ashtekar:1986yd}.

\acknowledgments

We are grateful to Jun'ichi Yokoyama for many helpful comments.
N. O. is supported by Grant-in-Aid for JSPS Fellow No.16J01780.
Y. P. W. is supported by Ministry of Science and Technology (MoST) Postdoctoral Research Abroad Program (PRAP)
MoST-105-2917-I-564-022.




\begin{thebibliography}{99}

\bibitem{Gibbons:1976ue} 
G.~W.~Gibbons and S.~W.~Hawking,
Phys.\ Rev.\ D {\bf 15}, 2752 (1977).
doi:10.1103/PhysRevD.15.2752

\bibitem{York:1972sj} 
J.~W.~York, Jr.,
Phys.\ Rev.\ Lett.\  {\bf 28}, 1082 (1972).
doi:10.1103/PhysRevLett.28.1082


\bibitem{Hawking:1978} 
S. W.~Hawking,
Phys. Rev. D  {\bf 18}, p. 1747-1753 (1978).



\bibitem{Einstein:1928}
A. Einstein, Sitz. Preuss. Akad. Wiss., 1928, p. 217; 
A. Einstein, Sitz. Preuss. Akad. Wiss., 1928, p. 224; 
A. Unzicker, T. Case, arXiv:physics/0503046.

\bibitem{Cartan}
E. Cartan, A. Einstein, Letters on Absolute Parallelism 1929-1932,
Princeton University Press, 1979. 

\bibitem{Weitzenbock:1923}
R.~Weitzenb\"{o}ck,  1923 Invariantentheorie (Gronningen:
Noordhoff)
	
\bibitem{Aldrovandi:2012}
R. Aldrovandi, J.G. Pereira, Teleparallel Gravity: An Introduction, Springer, Heidelberg, Germany, 2012.	

\bibitem{Einstein:1929}
A. Einstein,
Sitz. Preuss. Akad. Wiss., {\bf 18} (1929), p.156-159.


\bibitem{Cho:1975dh} 
Y.~M.~Cho,
Phys.\ Rev.\ D {\bf 14}, 2521 (1976).

\bibitem{Hayashi:1979qx} 
K.~Hayashi and T.~Shirafuji,
Phys.\ Rev.\ D {\bf 19}, 3524 (1979)
Addendum: [Phys.\ Rev.\ D {\bf 24}, 3312 (1982)].

\bibitem{Maluf:1994}
J.~Maluf,
J.\ Math.\ Phys. {\bf 35} (1994), 335.

\bibitem{Wu:2016dkt} 
Y.~P.~Wu,
Phys.\ Lett.\ B {\bf 762}, 157 (2016)
doi:10.1016/j.physletb.2016.09.025
[arXiv:1609.04959 [gr-qc]].

\bibitem{York:1986} 
J.~W.~York,
Found Phys (1986) 16: 249. 

\bibitem{Wu:2011kh} 
Y.~P.~Wu and C.~Q.~Geng,
Phys.\ Rev.\ D {\bf 86}, 104058 (2012)
doi:10.1103/PhysRevD.86.104058
[arXiv:1110.3099 [gr-qc]].

\bibitem{Arnowitt:1959ah} 
R.~L.~Arnowitt, S.~Deser and C.~W.~Misner,
Phys.\ Rev.\  {\bf 116}, 1322 (1959).

\bibitem{Oshita:2016oqn} 
N.~Oshita and J.~Yokoyama,
PTEP {\bf 2016}, no. 5, 053E02 (2016)
doi:10.1093/ptep/ptw053
[arXiv:1603.06671 [hep-th]].

\bibitem{Indices}
We use the lower case indices
$i, j, ...$ for the spatial coordinate and the lower case indices
$a, b, ...$ from the beginning of the alphabet are for the spatial
components of a local tangent frame.

\bibitem{Padmanabhan:2009vy} 
T.~Padmanabhan,
Rept.\ Prog.\ Phys.\  {\bf 73}, 046901 (2010)
[arXiv:0911.5004 [gr-qc]].
	
\bibitem{Hawking:1995fd} 
S.~W.~Hawking and G.~T.~Horowitz,
Class.\ Quant.\ Grav.\  {\bf 13}, 1487 (1996)
doi:10.1088/0264-9381/13/6/017
[gr-qc/9501014].

\bibitem{Bekenstein:1973ur} 
J.~D.~Bekenstein,
Phys.\ Rev.\ D {\bf 7}, 2333 (1973).
doi:10.1103/PhysRevD.7.2333
	
\bibitem{Krssak:2015rqa} 
M.~Kr\v{s}\v{s}\'{a}k and J.~G.~Pereira,
Eur.\ Phys.\ J.\ C {\bf 75}, no. 11, 519 (2015)
doi:10.1140/epjc/s10052-015-3749-2
[arXiv:1504.07683 [gr-qc]].
	
\bibitem{Krssak:2015lba} 
M.~Kr\v{s}\v{s}\'{a}k,
Eur.\ Phys.\ J.\ C {\bf 77}, no. 1, 44 (2017)
doi:10.1140/epjc/s10052-017-4621-3
[arXiv:1510.06676 [gr-qc]].

\bibitem{Geng:2011aj} 
C.~Q.~Geng, C.~C.~Lee, E.~N.~Saridakis and Y.~P.~Wu,
Phys.\ Lett.\ B {\bf 704}, 384 (2011)
[arXiv:1109.1092 [hep-th]].

\bibitem{Sotiriou:2010mv} 
T.~P.~Sotiriou, B.~Li and J.~D.~Barrow,
Phys.\ Rev.\ D {\bf 83}, 104030 (2011)
[arXiv:1012.4039 [gr-qc]].
	
\bibitem{Jarv:2015odu} 
L.~Jarv and A.~Toporensky,
Phys.\ Rev.\ D {\bf 93}, no. 2, 024051 (2016)
[arXiv:1511.03933 [gr-qc]].
	
\bibitem{Li:2010cg} 
However, generalized teleparallel theories beyond the GR limit lose local 
Lorentz invariance in the tagent frame, see
B.~Li, T.~P.~Sotiriou and J.~D.~Barrow,
Phys.\ Rev.\ D {\bf 83}, 064035 (2011)
[arXiv:1010.1041 [gr-qc]]; and Ref. \cite{Wu:2016dkt}. 
	
\bibitem{Dyer:2008hb} 
E.~Dyer and K.~Hinterbichler,
Phys.\ Rev.\ D {\bf 79}, 024028 (2009)
doi:10.1103/PhysRevD.79.024028
[arXiv:0809.4033 [gr-qc]].

\bibitem{Nelson:1977qj} 
J.~E.~Nelson and C.~Teitelboim,
Phys.\ Lett.\  {\bf 69B}, 81 (1977).
doi:10.1016/0370-2693(77)90138-1

\bibitem{Nelson:1978ex} 
J.~E.~Nelson and C.~Teitelboim,
Annals Phys.\  {\bf 116}, 86 (1978).
doi:10.1016/0003-4916(78)90005-2
	
\bibitem{Ashtekar:1986yd} 
A.~Ashtekar,
Phys.\ Rev.\ Lett.\  {\bf 57}, 2244 (1986).
doi:10.1103/PhysRevLett.57.2244
	
\end{thebibliography}
\end{document}